\documentstyle[12pt]{article}
\newcommand{\bea}{\begin{eqnarray}}
\newcommand{\eea}{\end{eqnarray}}
\newcommand{\be}{\begin{equation}}
\newcommand{\ee}{\end{equation}}
\newcommand{\p}{\prime}
\newcommand{\nn}{\nonumber}
\newcommand{\rf}[1]{(\ref{#1})}

\begin{document}

\begin{center}
{\Large\bf QUASIGROUP OF LOCAL-SYMMETRY TRANSFORMATIONS IN CONSTRAINED
THEORIES}\\
\vspace*{1cm}
{\large N.P.Chitaia, S.A.Gogilidze}\\
{\it Tbilisi State University, University St.9, 380086 Tbilisi, Georgia,}\\
\bigskip and
\\
\bigskip
{\large{Yu.S.Surovtsev}} \\
{\it Joint Institute for Nuclear Research, Dubna 141 980, Moscow Region,
Russia}
\begin{abstract}
In the framework of the generalized Hamiltonian formalism by Dirac, the local
symmetries of dynamical systems with first- and second-class constraints are
investigated in the general case without restrictions on the algebra of
constraints. The method of constructing the generator of local-symmetry
transformations is obtained from the requirement for them to map the solutions
of the Hamiltonian equations of motion into the solutions of the same
equations. It is proved that second-class constraints do not contribute to the
transformation law of the local symmetry entirely stipulated by all the
first-class constraints (and only by them) of an equivalent set passing to
which from the initial constraint set is always possible and is presented.
A mechanism of occurrence of higher derivatives of coordinates and group
parameters in the symmetry transformation law in the Noether second theorem
is elucidated. In the latter case it is shown that the obtained
transformations of symmetry are canonical in the extended (by Ostrogradsky)
phase space. It is thereby shown in the general case that the degeneracy of
theories with the first- and second-class constraints is due to their
invariance under local-symmetry transformations. It is also shown in the
general case that the action functional and the corresponding Hamiltonian
equations of motion are invariant under the same quasigroup of
local-symmetry transformations.
\end{abstract}
\end{center}
\section{Introduction}
In our previous paper \cite{CGS-2} (below cited as paper II) we have
considered constrained special-form theories with first- and second-class
constraints (when the first-class primary constraints are the ideal of a
quasi-algebra of all the first-class constraints) and have suggested the
method of constructing the generator of local-symmetry transformations in both
the phase and configuration space. It was proved that second-class
constraints do not contribute to the transformation law of the local symmetry
which entirely is stipulated by all the first-class constraints unlike the
assertions appeared recently in the literature
\cite{Sugano-Kimura}-\cite{Lusanna}. It was thereby shown that degeneracy of
special-form theories with the first- and second-class constraints is due to
their quasi-invariance under local-symmetry transformations. One must say the
mentioned restriction on an algebra of constraints is fulfilled in most of the
physically interesting theories, e.g., in electrodynamics, in the Yang --
Mills theories, in the Chern -- Simons theory, etc., and it has been used by
us in previous works \cite{GSST} in the case of dynamical systems only with
first-class constraints and also by other authors at obtaining gauge
transformations on the basis of different approaches
\cite{Bergmann}-\cite{Cabo}, \cite{Castellani,Cawley}. However, in the
existing literature there are examples of Lagrangians where this condition on
constraints does not hold, e.g., Polyakov's string \cite{Polyakov} and other
model Lagrangians \cite{Cawley}, \cite{Frenkel}-\cite{GSST:JP}. Then it was
natural to ask: Can the local-symmetry transformations be obtained in these
theories? What is a role of second-class constraints under these
transformations and, generally, what is the nature of the Lagrangian
degeneracy in this case? For example, in ref.\cite{Gracia-Pons} it is stated
that in the mentioned example the gauge transformation generators do not exist
for the Hamiltonian formalism though for the Lagrangian one the gauge
transformations may be constructed. In refs.\cite{GSST:tmf2,GSST:JP} in the
case of theories only with first-class constraints we have shown that one can
always pass to equivalent sets of constraints, for which the indicated
condition holds valid, and, therefore, gauge transformations do exist both in
the Hamiltonian and Lagrangian formalism. Therefore, the degeneracy of
theories with the first-class constraints is due to their invariance under
gauge transformations without restrictions on the algebra of constraints.

In the present paper it will be shown that, as in the presence only of
first-class constraints, in the general case of systems with first- and
second-class constraints, when the mentioned condition on constraints is not
fulfilled, there always exist equivalent sets of constraints, for which the
indicated condition holds valid. Therefore, the conclusions made in the
former case about the existence of local-symmetry transformations in both the
Hamiltonian and Lagrangian formalism and about the nature of degeneracy of
theories hold valid also in the general case. Also the conclusion of paper II
about the no influence of second-class constraints on local-symmetry
transformations and the conclusion of ref.\cite{GSST:tmf1} about the
mechanism of appearance of higher derivatives of coordinates and of group
parameters in these transformations are valid in the general case.

One can see that in the case, when higher (than first order) derivatives
of coordinates enter into the Noether transformation law in the configuration
space, the generator of local-symmetry transformations in the phase space
depends on derivatives of coordinates and momenta. Therefore, the Poisson
brackets are not determined in this case, and there arises a question about
the canonicity of the obtained transformations. Here we shall show that the
difficulty with the Poisson brackets is surmounted through the extension by
Ostrogradsky of phase space and the proof of canonicity of local-symmetry
transformations in this phase space, which had been furnished by us earlier
for theories only with first-class constraints \cite{GSST:tmf1}, hold true
also in the presence of second-class constraints in theory.

This paper is organized as follows. In section 2, for the general case of
systems with first- and second-class constraints (without restriction on the
algebra of first-class constraints) we derive the local-symmetry
transformations from the requirement for them to map the solutions of the
Hamiltonian equations of motion into the solutions of the same equations.
The derivation of a generator from this requirement (unlike the one from
quasi-invariance of the action functional in paper II) is made to establish a
ratio of the groups of local-symmetry transformations under which the
equations of motion and the action functional are invariant (as it is known,
generally, the action functional is invariant under a more slender group of
symmetry transformations than the corresponding equations of motion do). As
in paper II, these derivations are based substantially on results of our
previous paper \cite{CGS-1} (paper I) on the separation of constraints into
the first- and second-class ones and on properties of the canonical set of
constraints. In section 3, we consider the local-symmetry transformations in
the extended (by Ostrogradsky) phase space. In the 4th section the method is
illustrated by an example. In Appendix A, we describe the way of passing to an
equivalent constraint set when all the primary constraints of the first class
are momentum variables.

\section{Local-Symmetry Transformations in General Case of Systems with
First- and Second-Class Constraints}

As in the special case (paper II; below we shall refer to formulas of papers I
and II as (I.$\cdots$) and (II.$\cdots$)), we shall consider a dynamical
system with the canonical set $(\Phi_\alpha^{m_\alpha},\Psi_{a_i}^{m_{a_i}})$
of first- and second-class constraints, respectively $(\alpha=1,\cdots,F,~
m_\alpha=1,\cdots,M_\alpha;~~a_i=1,\cdots,A_i,~m_{a_i}=1,\cdots,M_{a_i},~i=1,
\cdots,n)$, properties of which are expressed by the Poisson brackets among
them and the Hamiltonian by the formulas (II.9)-(II.12):
\bea
&& \bigl\{\Phi_\alpha^{m_\alpha},H\bigr\} =
g_{\alpha~~\beta}^{m_\alpha m_\beta}~\Phi_\beta^{m_\beta},\quad
m_\beta=1,\cdots,m_\alpha+1,~~~~~~~~~~~~~~~~~~~ \label{PB-Phi-H^prime}\\
&& \bigl\{\Psi_{a_i}^{m_{a_i}},H\bigr\} = \bar{g}_{{a_i}~~\alpha}^{m_{a_i}
m_\alpha}~\Phi_\alpha^{m_\alpha}+\sum_{k=1}^{n}h_{{a_i}~~{b_k}}^{m_{a_i}
m_{b_k}}~\Psi_{b_k}^{m_{b_k}},\quad m_{b_n}=m_{a_i}+1,~~~~~~~\label{PB-Psi-H^prime}\\
&& \bigl\{\Phi_\alpha^{m_\alpha},\Phi_\beta^{m_\beta}\bigr\} =
f_{\alpha~~\beta~~\gamma}^{m_\alpha m_\beta m_\gamma}~\Phi_\gamma^{m_\gamma},
~~~~~~~~~~~~~~~~~~~\label{PB-Phi-Phi}\\
&& \bigl\{\Psi_{a_i}^{m_{a_i}},\Psi_{b_k}^{m_{b_k}}\bigr\} =\bar{f}_
{{a_i}~~{b_k}~~\gamma}^{m_{a_i} m_{b_k} m_\gamma}~\Phi_\gamma^{m_\gamma}+
\sum_{l=1}^{n}k_{{a_i}~~{b_k}~~{c_l}}^{m_{a_i} m_{b_k} m_{c_l}}~\Psi_{c_l}^
{m_{c_l}}+D_{{a_i}~~{b_k}}^{m_{a_i} m_{b_k}}~~~~~~~~ \label{PB-Psi-Psi}
\eea
with general properties of the structure functions given by the formulas
(II.13)-(II.16)
\be \label{g}
g_{\alpha~~\beta}^{m_\alpha m_\beta}=0,\quad\mbox{if}~m_\alpha+2\leq m_\beta~,
\ee
\bea \label{bar-g-h}
\left\{\begin{array}{ll}
\bar{g}_{{a_i}~~\alpha}^{m_{a_i} m_\alpha}=0,&~~\mbox{if}~~ m_\alpha\geq
m_{a_i}~,\\{}\\ h_{{a_i}~~{b_k}}^{m_{a_i} m_{b_k}}=0,&~~\mbox{if}~~ m_{a_i}+2
\leq m_{b_k}~~\mbox{or if}~~a_i=b_k,~~m_{a_i}=M_{a_i},\\
{} &~~m_{b_k}\geq M_{a_i}, \end{array}\right.
\eea
\be \label{f}
f_{\alpha~~\beta~~\gamma}^{1~ m_\beta m_\gamma}=0~~~\mbox{for}~~
m_\gamma\geq 2,
\ee
\bea \label{F-sym}
\left\{\begin{array}{l}
F_{{a_i}~~~~~{b_i}}^{M_{a_i}-l~l+1}=(-1)^l~F_{{a_i}~{b_i}}^{1~M_{b_i}}~,~~
~~l=0,1,\cdots,M_{a_i}-1,\\ {}\\
F_{{a_i}~{b_i}}^{j~k}=0,~~~~~~~\mbox{if}~~j+k\neq M_{a_i}+1,\\ {}\\
F_{{a_i}~~{b_k}}^{m_{a_i} m_{b_k}}=0,~~~~\mbox{if}~~a_i,b_k~\mbox
{refer to different chains (or doubled}\\ ~~~~~~~~~~~~\mbox{chains) of
second-class constraints}~
\bigl(D_{{a_i}~~{b_k}}^{m_{a_i} m_{b_k}}~\stackrel{\Sigma}{=}F_{{a_i}~~{b_k}}^
{m_{a_i} m_{b_k}}\bigr)
\end{array}\right.
\eea
and with
\be \label{H}
H = H_c+\sum_{k=1}^{n}({\bf K}^{1~k})_{b_k~a_k}^{-1}\{\Psi_{a_k}^k,H_c\}
\Psi_{b_k}^1
\ee
being a first-class function \cite{Dirac}; $H_c$ is the canonical
Hamiltonian.\\
Passing to this set from the initial one is always possible in an arbitrary
case by the method developed in paper I. Here we shall consider the general
case (when first-class primary constraints are not the ideal of quasi-algebra
of all the first-class constraints, i.e. the restriction (II.25) is not
fulfilled) and derive local-symmetry transformations.

A group of phase-space coordinate transformations that maps each solution of
the Hamiltonian equations of motion into the solution of the same equations
will be called the symmetry transformation.

Consider the Hamiltonian equations of motion in the following form:
\bea \label{mot.eq-ns}
\left\{\begin{array}{l}
\dot q_i~\stackrel{\Sigma_1}{\approx}~\{q_i,H_T\},\quad \dot p_i ~\stackrel
{\Sigma_1}{\approx}~ \{p_i,H_T\}, \quad i=1,\cdots,N,\\ 
\Psi_{a_k}^1~\stackrel{\Sigma_1}{\approx}0,~~a_k=1,\cdots,A_k~(k=1,\cdots,n),\\
\Phi_\alpha^1~\stackrel{\Sigma_1}{\approx}0,~~ \alpha=1,\cdots,F,
\end{array}\right.
\eea
where
\be \label{H_T}
H_T = H + u_\alpha\Phi_\alpha^1,
\ee
$u_\alpha$ are undetermined Lagrange multipliers; the symbol
$\stackrel{\Sigma_1} {\approx}$ means weak equality on the primary-constraints
surface $\Sigma_1$.

Consider also the infinitesimal transformations of the phase-space coordinates
\bea \label{q-q^prime}
\left\{\begin{array}{ll} q_i^\p = q_i+\delta q_i ,& \quad
\delta q_i = \{q_i , G\} ,
\\ p_i^\p = p_i+\delta p_i ,& \quad\delta p_i =
\{p_i , G\} \end{array}\right.
\eea
with the generator $G$ sought in the form (II.4)
\be \label{G-Dirac}
G=\varepsilon_\alpha^{m_\alpha}\Phi_\alpha^{m_\alpha} +
\eta_{a_i}^{m_{a_i}}\Psi_{a_i}^{m_{a_i}}.
\ee
To recognize a role of the second-class constraints in the local-symmetry
transformations in this general case, we consider them on the same basis as
the first-class constraints.

Like in refs.\cite{Cabo,Castellani,Gracia-Pons,GSST:JP}, we will require
the transformed quantities $q_i^\p(t)$ and $p_i^\p(t)$ defined by
\rf{q-q^prime} to be solutions of the Hamiltonian equations of motion
\rf{mot.eq-ns} provided that the initial $q_i(t)$ and $p_i(t)$ do this, i.e.
\bea \label{mot.eq-ns'prime} &&\dot
q_i^\p~\stackrel{\Sigma_1}{\approx}~\frac{\partial H_T^\p}{q_i}(q^\p,
p^\p),\quad \dot p_i^\p ~\stackrel {\Sigma_1}{\approx}~\frac{\partial H_T^\p}
{p_i}(q^\p,p^\p), \quad i=1,\cdots,N,\nn\\ 
&&\Psi_{a_k}^1(q^\p,p^\p)~\stackrel{\Sigma_1}{\approx}0,~~a_k=1,\cdots,A_k~(k=1,
\cdots,n),\\
&&\Phi_\alpha^1(q^\p,p^\p)~\stackrel{\Sigma_1}{\approx}0,~~ \alpha=1,\cdots,F,
\nn
\eea
where
\be \label{H_T'prime}
H_T^\p=H_T+\delta u_\alpha(t)\Phi_\alpha^1(q,p)=H+u_\alpha^\p(t)\Phi_\alpha^1
(q,p).
\ee
Replacements in \rf{H_T'prime} of $H_T$ by $H$ and of $u_\alpha(t)$ by
$u_\alpha^\p(t)$ are stipulated by that, generally speaking, different
solutions that should be related with each other through the local-symmetry
transformations correspond to different choices of the functions $u_\alpha(t)$
(the transformed quantities are denoted by the same letters with the prime).
In equations \rf{mot.eq-ns'prime} it is taken into consideration that the
transformations \rf{q-q^prime} must conserve the primary-constraints surface
$\Sigma_1$ (see the argument after formula (6) in paper II).

Equations \rf{mot.eq-ns'prime} can be rewritten with taking account of
\rf{q-q^prime} and \rf{mot.eq-ns} in the following form:
\bea
&&\frac{d}{dt}\{q_i,G\}~\stackrel{\Sigma_1}{\approx}~\{\{q_i,H_T^\p\},G\},
\label{G:mot.eq-ns-q}\\
&&\frac{d}{dt}\{p_i,G\}~\stackrel{\Sigma_1}{\approx}~\{\{p_i,H_T^\p\},G\},
\quad i=1,\cdots,N,\label{G:mot.eq-ns-p}\\ 
&&\bigl\{\Psi_{a_k}^1,G\bigr\}~\stackrel{\Sigma_1}{\approx}0,~~a_k=1,\cdots,
A_k~(k=1,\cdots,n),\label{PB-G-Psi1}\\
&&\bigl\{\Phi_\alpha^1,G\bigr\}~\stackrel{\Sigma_1}{\approx}0,~~ \alpha=1,
\cdots,F. \label{PB-G-Phi1}
\eea
We shall analyze consequences of the obtained equation system starting from
the conditions of the primary-constraints surface conservation \rf{PB-G-Psi1}
and \rf{PB-G-Phi1}. As in the special case of paper II (the consideration
is completely identical), from \rf{PB-G-Psi1} we obtain that in expression
\rf{G-Dirac} the coefficients of those $i$-ary constraints, which are the
final stage of each chain of second-class constraints, and of those
second-class primary constraints, which do not generate the secondary
constraints, disappear:
\be \label{eta-i=0}
\eta_{a_i}^i=0\quad \mbox{for}~ i=1,\cdots,n,
\ee
As to the condition \rf{PB-G-Phi1}, we rewrite it in the form:
\be \label{PB-G-Phi1-detail}
\bigl\{\Phi_\alpha^1,G\bigr\}=\bigl(f_{\alpha~~\beta~~\gamma}^{1~ m_\beta~1}~
\Phi_\gamma^1+f_{\alpha~~\beta~~\gamma}^{1~~m_\beta~m_\gamma}~\Phi_\gamma^
{m_\gamma}\bigr)\varepsilon_\beta^{m_\beta}+\bigl\{\Phi_\alpha^1,\Psi_{a_i}^
{m_{a_i}}\bigr\}\eta_{a_i}^{m_{a_i}}~\stackrel{\Sigma_1}{\approx}0,~
\ee
$$\alpha,\beta,\gamma=1, \cdots,F;~~m_\beta=1, \cdots,M_\beta;~~
m_\gamma=2, \cdots,M_\gamma.~~~~~~$$
The last term in \rf{PB-G-Phi1-detail} vanish for the canonical set of
constraints $(\Phi,\Psi)$; therefore, the equality \rf{PB-G-Phi1-detail} were
satisfied if ~$f_{\alpha~~\beta~~\gamma}^{1~~m_\beta~m_\gamma}=0$~ for
$m_\gamma\geq 2$~ (i.e. the first-class primary constraints were the ideal of
quasi-algebra of all the first-class constraints). This case is considered in
paper II. Here we consider the general case of a constraint algebra when
\be \label{f-gen}
f_{\alpha~~\beta~~\gamma}^{1~~m_\beta~m_\gamma}\neq 0 \quad\mbox{for}\quad
m_\gamma \geq 2. \ee
For systems only with first-class constraints, the case \rf{f-gen} was
investigated by us earlier \cite{GSST:tmf2,GSST:JP}. For systems with first-
and second-class constraints, when \rf{f-gen} is the case, one can act in the
same way as in the presence only of first-class constraints, i.e. using
arbitrariness that is inherent in the generalized Hamiltonian formalism by
Dirac, we shall pass to an equivalent set of constraints by the transformation
that affects only first-class constraints:
\be \label{Phi-bar-Phi}
\tilde{\Phi}_\beta^{m_\beta}=C_{\beta~~\alpha}^{m_\beta m_\alpha}\Phi_\alpha^
{m_\alpha},\qquad \mbox{det}\left\|C_{\beta~~\alpha}^{m_\beta
m_\alpha}\right\|_\Sigma \not =0.
\ee
It is sufficient to consider a particular case of the transformation
\rf{Phi-bar-Phi} when primary constraints remain unchanged, i.e.
$$C_{\beta~~\alpha}^{1~~m_\alpha}=\delta_{\beta\alpha}\qquad\mbox{ for any }
\quad m_\alpha.$$
It is not difficult to see that taking account of \rf{PB-Phi-Phi} we obtain
\bea \label{PB-Phi1-bar-Phi}
\bigl\{\Phi_\alpha^1,\tilde{\Phi}_\beta^{m_\beta}\bigr\}&=&\bigl[\bigl\{\Phi_
\alpha^1,C_{\beta~~\gamma}^{m_\beta m_\gamma}\bigr\}+f_{\alpha~~\delta~~
\gamma}^{1~~m_\delta m_\gamma}C_{\beta~~\delta}^{m_\beta m_\delta}\bigr]\Phi_
\gamma^{m_\gamma} \nn\\
&&~+f_{\alpha~~\delta~~\gamma}^{1~~m_\delta~1}C_{\beta~~\delta}^
{m_\beta m_\delta}\Phi_\beta^1,\qquad m_\beta,m_\delta,m_\gamma\geq 2.
\eea
>From the expression \rf{PB-Phi1-bar-Phi} it is clear that if we could choose
$C_{\beta~~\gamma}^{m_\beta m_\gamma}$ so that the coefficients of secondary
constraints vanish
\be \label{C-equation}
\bigl\{\Phi_\alpha^1,C_{\beta~~\gamma}^{m_\beta m_\gamma}\bigr\}+f_{\alpha~~
\delta~~\gamma}^{1~~m_\delta m_\gamma}C_{\beta~~\delta}^{m_\beta m_\delta}=0,
\ee
for a new set of constraints $\tilde{\Phi}_\beta^{m_\beta}$ we were obtained
~$\tilde{f}_{\alpha~~\beta~~\gamma}^{1~~m_\beta~m_\gamma}=0$~ (for $m_\gamma
\geq 2$) and
\be \label{ideal}
\bigl\{\tilde{\Phi}_\alpha^1,\tilde{\Phi}_\beta^{m_\beta}\bigr\}=
\tilde{f}_{\alpha~~\beta~~\gamma}^{1~~m_\delta~1}~\tilde{\Phi}_\gamma^1,
\ee
i.e. that is needed for the realization of \rf{PB-G-Phi1-detail}. Thus, for
$C_{\beta~~\gamma}^{m_\beta m_\gamma}$ we have derived the system of linear
inhomogeneous equations in the first-order partial derivatives
\rf{C-equation}. This system can be shown to be fully integrable. The
condition of integrability for systems of the type \rf{C-equation} looks as
follows \cite{Smirnov}
\be \label{C:integr.cond.}
\bigl\{\Phi_\sigma^1,\bigl\{\Phi_\alpha^1,C_{\beta~~\gamma}^{m_\beta m_\gamma}
\bigr\}\bigr\}-\bigl\{\Phi_\alpha^1,\bigl\{\Phi_\sigma^1,C_{\beta~~\gamma}^
{m_\beta m_\gamma}\bigr\}\bigr\}=0.
\ee
Using eq.\rf{C-equation}, properties of the Poisson brackets and making some
transformations we rewrite the relation \rf{C:integr.cond.} in the form
\bea \label{Phi-f-C}
& & \bigl[\bigl\{\Phi_\alpha^1,f_{\sigma~~\delta~~\gamma}^{1~~m_\delta m_
\gamma}\bigr\}-
f_{\alpha~~\delta~~\tau}^{1~~m_\delta m_\tau}f_{\sigma~~\tau~~\gamma}^{1~~
m_\tau m_\gamma}-\bigl\{\Phi_\sigma^1,f_{\alpha~~\delta~~\gamma}^{1~~m_\delta
m_\gamma}\bigr\} \qquad\nn\\
& &~ +f_{\sigma~~\delta~~\tau}^{1~~m_\delta m_\tau}f_{\alpha~~\tau~~\gamma}^
{1~~m_\tau m_\gamma}\bigr]C_{\beta~~\delta}^{m_\beta m_\delta}=0,
\qquad m_\beta, m_\delta ,m_\tau\geq 2.
\eea
Utilizing the Jacobi identity
$$\bigl\{\Phi_\alpha^1,\bigl\{\Phi_\sigma^1,\Phi_\beta^{m_\beta}\bigl\}\bigr\}
+ \bigl\{\Phi_\beta^{m_\beta},\{\Phi_\alpha^1,\Phi_\sigma^1\}\bigr\} +
\bigl\{\Phi_\sigma^1,\{\Phi_\beta^{m_\beta},\Phi_\alpha^1\}\bigr\} = 0,
\qquad m_\beta\geq 2 $$
and the relation \rf{PB-Phi-Phi} we obtain
\bea \label{identity:Phi-f-C}
& & \bigl[\bigl\{\Phi_\alpha^1,f_{\sigma~~\delta~~\gamma}^{1~~m_\delta
m_\gamma}\bigr\}- f_{\alpha~~\delta~~\tau}^{1~~m_\delta
m_\tau}f_{\sigma~~\tau~~\gamma}^{1~~ m_\tau
m_\gamma}-\bigl\{\Phi_\sigma^1,f_{\alpha~~\delta~~\gamma}^{1~~m_\delta
m_\gamma}\bigr\} \nn\\ & &~+ f_{\sigma~~\delta~~\tau}^{1~~m_\delta
m_\tau}f_{\alpha~~\tau~~\gamma}^ {1~~m_\tau
m_\gamma}\bigr]\Phi_\gamma^{m_\gamma} =
\bigl\{\bigl\{\Phi_\alpha^1,\Phi_\sigma^1\bigr\},\Phi_\delta^{m_\delta}\bigr\},
\eea
$$ m_\beta\geq 2,\quad m_\gamma, m_\delta ,m_\tau\geq 1.$$
Note that every primary constraint of first class contains at least one
momentum variable, therefore, there always exist canonical transformations
transforming the primary constraints into new momentum variables (see Appendix
A). We shall regard such transformation to be carried out, therefore, the
Poisson brackets between primary constraints may be considered to be strictly
zero in the whole phase space. From here, the expressions in the square
brackets in front of the constraints $\Phi_\gamma^{m_\gamma}$ on the left-hand
side of the identity \rf{identity:Phi-f-C} being coefficients of the
functionally independent quantities disappear each separately. As the
condition \rf{Phi-f-C} contains the same coefficients of
$C_{\beta~~\delta}^{m_\beta m_\delta}$, it is satisfied identically, which
proves the system of equations \rf{C-equation} to be fully integrable.
Therefore, there always exists a set of constraints
$\tilde{\Phi}_\alpha^{m_\alpha}$ equivalent to the initial set for which the
condition \rf{ideal} (and, therefore, \rf{PB-G-Phi1}) holds valid. We shall
below omit the mark ``$\>\tilde{\;}\>$''.

Now, using the equality
\be \label{eq-ty:AB}
\frac{d}{dt}\bigl\{A,B\bigr\}=\bigl\{\frac{\partial A}{\partial t},B\bigr\}+
\bigl\{A,\frac{\partial B}{\partial t}\bigr\}+ \bigl\{\bigl\{A,B\bigr\},
H_T\bigr\}
\ee
(valid for arbitrary functions $A(q,p,t)$ and $B(q,p,t)$ given in the whole
phase space) and the Jacobi identities for the quantities $(q_i,G,H_T^\p)$ and
$(p_i,G,H_T^\p)$, we represent equations \rf{G:mot.eq-ns-q} and
\rf{G:mot.eq-ns-p} as
\bea
& & \bigl\{q_i,\frac{\partial G}{\partial t}+\bigl\{G,H_T^\p\bigr\}\bigr\}
\stackrel{\Sigma_1}{\approx} 0, \label{eq:q-G} \\
& & \bigl\{p_i,\frac{\partial G}{\partial t}+\bigl\{G,H_T^\p\bigr\}\bigr\}
\stackrel{\Sigma_1}{\approx} 0, \label{eq:p-G}
\eea
respectively. By virtue of an arbitrariness of the multipliers $u_\alpha(t)$,
in what follows the prime will be omitted. If these equalities were the
case in the whole phase space, it would follow from them that
$$\frac{\partial G(q,p,t)}{\partial t}+\bigl\{G(q,p,t),H_T(q,p,t)\bigr\}=
f(t),$$
where $f(t)$ is an arbitrary function of time. However, since eqs.\rf{eq:q-G}
and \rf{eq:p-G} are the case only on the surface $\Sigma_1$, we obtain that
\be \label{eq:G}
\frac{\partial G(q,p,t)}{\partial t}+\bigl\{G(q,p,t),H_T(q,p,t)\bigr\}=
f(t)+J(q,p,t),
\ee
where
\bea
&&J=c_\alpha(q,p,t)\Phi_\alpha^1(q,p)+d_{a_i}(q,p,t)\Psi_{a_i}^1(q,p),\nn\\
&&\alpha=1,\cdots,F,~~~~a_i=1,\cdots,A_i,~~i=1,\cdots,n.\nn
\eea
However, both $f(t)$ and $J(q,p,t)$ are identity generators on the primary
constraint surface, and can be ignored in subsequent discussions
\cite{Castellani}. Note that equation \rf{eq:G} (with $f(t)$ ignored) is a
necessary condition of that $G$ is the generating function of infinitesimal
transformations of local symmetry \rf{q-q^prime}, and, furthermore, this is
sufficient for a quasi-invariance (within a surface term) of the action
functional
\be \label{S}
S=\int_{t_1}^{t_2}dt~(p\dot q - H_T),
\ee
under these transformations. To see the latter, consider the variation of
action, induced by the transformations \rf{q-q^prime},
$$ \delta S=\int_{t_1}^{t_2}dt\Bigl[\frac{d}{dt}(p_i\frac{\partial G}
{\partial p_i}-\frac{\partial G}{\partial q_i}\dot q_i-\frac{\partial G}
{\partial p_i}\dot p_i+\{G,H_T\}\Bigr] $$
which, with taking into account the relation
$$\frac{dG}{dt}=\frac{\partial G}{\partial t}+\frac{\partial G}{\partial q_i}
\dot q_i+\frac{\partial G}{\partial p_i}\dot p_i, $$
can be rewritten as
\be \label{delta-S}
\delta S=\int_{t_1}^{t_2}dt\Bigl[\frac{d}{dt}(p_i\frac{\partial G}
{\partial p_i}-G)+\frac{\partial G}{\partial t}+\{G,H_T\}\Bigr]
\ee
giving the desired result if eq.\rf{eq:G} is fulfilled.

Now, inserting the required form of the generator $G$ \rf{G-Dirac} into
\rf{eq:G}, we obtain the equality (II.17) which must be satisfied by a proper
inspection of the coefficients $\varepsilon_\alpha^{m_\alpha}$ and
$\eta_{a_i}^{m_{a_i}}$. Further consideration repeats entirely the one of
paper II resulting in that the second-class constraints do not contribute
to the generator of local-symmetry transformations that is a linear
combination of all the first-class constraints (and only of them)
\be \label{G}
G=B_{\alpha~~\beta}^{m_\alpha m_\beta}\phi_\alpha^{m_\alpha}\varepsilon_\beta^
{(M_\alpha-m_\beta)} , \qquad m_\beta=m_\alpha,\cdots,M_\alpha .
\ee
with the coefficients
$$
B_{\alpha~~\beta}^{m_\alpha m_\beta}\varepsilon_\beta^{(M_\alpha-m_\beta)}
\quad \Bigl(~\varepsilon_\beta^{(M_\alpha-m_\beta)}\equiv
{{d^{M_\alpha-m_\beta}}\over{dt^{M_\alpha-m_\beta}}}\varepsilon_\beta (t),~~~
\varepsilon_\beta (t)\equiv \varepsilon_\beta^{M_\beta}~\Bigr) $$
determined from the system of equations
\be \label{eq:eps}
\dot \varepsilon_\alpha^{m_\alpha}+\varepsilon_\beta^{m_\beta}
g_{\beta~~\alpha}^{m_\beta m_\alpha}=0,\qquad m_\beta= m_\alpha-1,\cdots,
M_\alpha,
\ee
with the help of the procedure of reparametrization described in paper II.
The local-symmetry transformations of $q$ and $p$ determined by formulas
\rf{q-q^prime} are also the quasi-invariance transformations of the action
functional \rf{S}.

The corresponding transformations of local symmetry in the Lagrangian
formalism are determined in the following way:
\be \label{q-dot-q}
\delta q_i(t)=\{q_i(t),G\}\biggr|_{p=\frac{\partial L} {\partial \dot q}},
\qquad \delta\dot q(t) = \frac{d}{dt}\delta q(t).
\ee

So, one can state that in the general case of theories with first- and
second-class constraints (without restrictions on the constraint algebra) the
representation of a certain quantity $G$ as a linear combination of all the
first-class constraints (and only of them) with the coefficients determined by
the system of equations \rf{eq:eps} is the necessary and sufficient condition
for $G$ to be the local-symmetry transformation generator. In addition, these
are the necessary and sufficient conditions for \rf{q-q^prime} to be the
quasi-invariance transformation of the functional of action in both the phase
and $(q,\dot q)$ space.

\section{Local-Symmetry Transformations in the Extended Phase Space}

One can see that in the case, when higher (than first order) derivatives
of coordinates enter into the transformation law in the configuration spase
and into the surface term in the action variation, the coefficients
$B_{\alpha~~\beta}^{m_\alpha m_\beta}$ in expression \rf{G} for $G$ depend on
the derivatives of $q$ and $p$. It is clear, in this case there arises a
question about ``explicit'' canonicity of the obtained transformations outside
of the constraints surface. Therefore, it is clear that in the general case
one should consider not only the violation of the condition \rf{ideal}
(the manner of the deed in this case is worked out in the previous section)
but also that structure of constraints when there arise higher derivatives of
coordinates in the law of local-symmetry transformations. Here we shall show
how to construct these transformations in the latter case and prove the
canonicity of gauge transformations in the extended (by Ostrogradsky) phase
space, which has been shown by us earlier for theories with first-class
constraints \cite{GSST:tmf1}, to hold true also in the presence of
second-class constraints in a theory.

Let us consider the singular Lagrangian $L(q,\dot q)$, and let the higher
(than first) derivatives of coordinates contribute to the corresponding law of
local-symmetry transformations. Under these transformations we have
\be \label{L-prime}
L^\prime =L(q ,\dot q) + {d\over dt}F(q ,\dot q,\ddot q ,\cdots, \varepsilon,
\dot \varepsilon ,\cdots)
\ee
where $\varepsilon(t)$ are the group parameters. Adding to Lagrangian
$L(q,\dot q)$ the total time derivative of function which depends also on
higher derivatives does not change the Lagrangian equations of motion. As it
is seen from \rf{L-prime}, the theory with Lagrangian $L^\p$ must be
considered as the one with higher derivatives. Both Lagrangian and Hamiltonian
formulations of the theories with $L$ and $L^\p$ are equivalent \cite{Gitman}.
The Hamiltonian formulation of the theory with $L^\p$ is built in the extended
(by Ostrogradsky) phase space. An equivalence of Hamiltonian formulations of
the theories with $L$ and $L^\p$ means that the Hamiltonian equations of
motion of these both theories are related among themselves by canonical
transformations. Therefore, the Hamiltonian formulation of the theory with the
Lagrangian $L$ must be built in the same extended phase space as it is the
case for $L^\p$. Thus, the theory with $L$ will be considered from the very
beginning as the one with higher derivatives of the same order that they have
in $L^\p$.

>From the above reasoning it is clear that to require a canonicity of the
local-simmetry transformations has the meaning only in the indicated extended
phase space.

Let us construct the extended phase space using the formalism of theories with
higher derivatives \cite{Ostr,Gitman,Nesterenko}. We shall determine the
coordinates as follows
\be \label{enlarged-q}
q_{1~i}=q_i ,\quad q_{s~i}={d^{s-1}\over{dt^{s-1}}}~q_i ,\quad s=2 ,\cdots, K ,
\quad i=1 ,\cdots, N
\ee
where $K$ equals the highest order of derivatives of $q$ and $p$. The
conjugate momenta defined by the formula \cite{Ostr,Gitman,Nesterenko}
$$p_{r~i}=\sum_{l=r}^{K}(-1)^{l-r}\frac{d^{l-r}}{dt^{l-r}}\frac{\partial L}
{\partial q_{r+1~i}}$$
are
\be \label{enlarged-p}
p_{1~i}=p_i ,\qquad p_{s~i}=0\quad\mbox{ for }\quad s=2 ,\cdots, K.
\ee
The generalized momenta for $s\geq2$ are extra primary constraints of the
first class.

In the extended phase space the total Hamiltonian is written down as
\be \label{enlarged-H}
\bar H_T = H_T(q_{1~i} , p_{1~i}) + \lambda_{s~i}~p_{s~i} ,\quad s\geq 2,
\ee
where $H_T$ is of the same form as in the initial phase space \rf{H_T} and
$\lambda_{s~i}$ are arbitrary functions of time.

Now the Poisson brackets are determined in the following way
$$\bigl\{A,B\bigr\}=\frac{\partial A} {\partial q_{r~i}}\frac{\partial B}
{\partial p_{r~i}}-\frac{\partial A}{\partial p_{r~i}}\frac{\partial B}
{\partial q_{r~i}}.$$

>From \rf{enlarged-H} we may conclude that there do not appear additional
secondary constraints corresponding to $p_{s~i}$ for $s\geq 2$. The set of
constraints in the extended phase space remains the same as in the initial
phase space, obeys the same algebra \rf{PB-Phi-H^prime}-\rf{PB-Psi-Psi}, and
does not depend on the new coordinates and momenta as also $H_T$ does.

We shall seek a generator $\overline{G}$ in the extended phase space in
the form, analogous to the one in the initial phase space \rf{G-Dirac}. Then
from the requirements of quasi-invariance of the action
\be \label{enlarged-S}
\overline{S}=\int_{t_1}^{t_2}dt\bigl[p_{r~i}~q_{r+1~i} + p_{K~i}~\dot q_{K~i}
- \bar H_T \bigr] , \qquad r = 1 ,\cdots, K-1
\ee
and of conservation of the primary constraint surface ${\overline{\Sigma}}_1$
under the transformations generated by $\overline{G}$, we shall obtain the
same relations \rf{eq:eps} for determining $\varepsilon_\alpha^{m_\alpha}$
(with the help of the iterative procedure described in detail in  paper II)
and the same conclusion about no influence of second-class constraints on the
local symmetries of a system.

Before to implement the above-mentioned iterative procedure that gives the
result \rf{G}, we notice that the coefficients $B_{\alpha~~\beta}^{m_\alpha
m_\beta}$ would depend only on $q_{1~i}$ and $p_{1~i}$ and on their
derivatives. Now, carrying out the iterative procedure we shall exchange
derivatives of $q_{1~i}$ according to formula \rf{enlarged-q}, and for
derivatives of $p_{1~i}$ we shall make the following replacements:
\bea \label{dot-p}
p_{1~i} & = & \frac{\partial L}{\partial q_{2~i}} = h_0^i (q_{1~k} , q_{2~k}) ,
\qquad i,k=1,\cdots,N,\nn\\
\dot p_{1~i} & = & \frac{\partial h_0^i}{\partial q_{1~n}} q_{2~n} +
\frac{\partial h_0^i}{\partial q_{2~n}} q_{3~n} = h_1^i(q_{1~k} , q_{2~k} ,
q_{3~k}) ,\\ & \vdots & \nn\\
p_{1~i}^{(M_\alpha - 2)} & = & h_{M_\alpha - 2}^i (q_{1~k} , q_{2~k} ,\cdots,
q_{M_\alpha-1~k}) .\nn
\eea
As a result, we shall obtain the expression for $\overline{G}$:
\be \label{enl.G}
\overline{G}=B_{\alpha~~\beta}^{m_\alpha m_\beta}\Phi_\alpha^{m_\alpha}
\varepsilon_\beta^{(M_\alpha-m_\beta)} + \varepsilon_{s~i}p_{s~i},
\ee
$$m_\beta=m_\alpha,\cdots,M_\alpha,~~~  s=2 ,\cdots, K, $$
where $B_{\alpha~~\beta}^{m_\alpha m_\beta}(q_{1~i},\cdots,q_{M_\alpha -1~i};
p_{1~i})$, being just in the same forms as in the initial phase space, are
written, however, with taking account of the above-indicated replacements;
$\varepsilon_{s~i}$ are the supplementary group parameters in the amount
equal to the number of the supplementary primary constraints of first class
$p_{s~i}$. Note that the obtained generator \rf{enl.G} satisfies the group
property
\be \label{enl.G:group}
\{\overline{G}_1,\overline{G}_2\}=\overline{G}_3,
\ee
where the transformation $\overline{G}_3$ \rf{enl.G} is realized by carrying
out two successive transformations $\overline{G}_1$ and $\overline{G}_2$
\rf{enl.G}. Now the local-symmetry transformations of the coordinates of the
initial phase space in the extended one are of the form
\bea
\label{enl.-delta-q-p}
\left\{\begin{array}{l}
\delta q_{1~k}=\varepsilon_\beta^{(M_\alpha-m_\beta)}
\bigl\{q_{1~k},B_{\alpha~~\beta}^{m_\alpha m_\beta}(q_{1~i},\cdots,
q_{M_\alpha -1~i} ; p_{1~i})\phi_\alpha^{m_\alpha}(q_{1~i},p_{1~i})\bigr\},\\
{}\\
\delta p_{1~k}=\varepsilon_\beta^{(M_\alpha-m_\beta)}\bigl\{p_{1~k},
B_{\alpha~~\beta}^{m_\alpha m_\beta}(q_{1~i},\cdots,q_{M_\alpha -1~i} ;
p_{1~i})\phi_\alpha^{m_\alpha}(q_{1~i},p_{1~i})\bigr\}.
\end{array}\right.
\eea
One can verify that to within quadratic terms in $\delta q_{i~k}$ and
$\delta p_{j~n}$
$$\{q_{i~k}+\delta q_{i~k} ,p_{j~n}+\delta p_{j~n}\}=\delta_{ij}\delta_{kn},$$
i.e. the obtained infinitesimal transformations of local symmetry are
canonical in the extended (by Ostrogradsky) phase space.

The local-symmetry transformations in the configuration space may be obtained
if after calculating the Poisson brackets in the first formula
\rf{enl.-delta-q-p} one takes account of the definitions \rf{enlarged-q} and
of the generalized momenta $p_i$ and make use of formula \rf{q-dot-q} for
$\delta\dot q$. They are the Noether transformations. (Note that, as it is
seen from \rf{enl.-delta-q-p}, to reduce calculations in obtaining these
transformations one may use formulas \rf{q-q^prime} in the initial phase space
provided one applies the following ``rule'': derivatives of $q$ and $p$ are
simply put outside the Poisson brackets.) In this case, if the coefficients
$B_{\alpha~~\beta}^{m_\alpha m_\beta}$ depend explicitly on $q_{s~i}$, where
$s\geq 2$, then higher derivatives of coordinates $q_i^{(s)}(s\geq 2)$ are
present in the transformation law in the configuration space. The functions
$g_{\sigma~~\tau}^{m_\sigma m_\tau}$, arising in formula \rf{PB-Phi-H^prime},
signal to the appearance of that dependence. Moreover, the order of the
highest derivative of coordinates may be established already at the beginning,
when obtaining the explicit form of $g_{\sigma~~\tau}^{m_\sigma m_\tau}.$ To
this end, one ought to consider the systems of relations \rf{PB-Phi-H^prime}
and \rf{eq:eps} in a expanded form. One can see that if any of the coefficients
$g_{\alpha~~~~~\beta}^{{M_\alpha-1}~M_\alpha}$ and
$g_{\alpha~~~\beta}^{M_\alpha~M_\alpha}$ in front of the constraints of the
last stage $M_\alpha$ depends on $q_{1~i}$ and $p_{1~i}$, the coefficients
$B_{\alpha~~\beta}^{m_\alpha m_\beta}$ will depend on $q_{s~i}(s=2,\cdots,
M_\alpha-1)$, and the generator $\overline{G}$ will contain $q_{s~i}~(s=2,
\cdots, K)$, as it is seen from \rf{eq:eps}. Then, taking account of
\rf{enlarged-q}, the order of the highest possible derivative of coordinates
in the law of the Noether transformations in the configuration space is equal
to $K\equiv\mbox{max}_{\alpha}(M_\alpha-1)$. If these coefficients are
constants and any of coefficients
$g_{\alpha~~~~~\beta}^{{M_\alpha-2}~{M_\alpha-1}}$,
$g_{\alpha~~~~~\beta}^{{M_\alpha-1}~{M_\alpha-1}}$ and
$g_{\alpha~~~\beta}^{M_\alpha~{M_\alpha-1}}$
in front of the constraints of the antecedent stage $\phi_\beta^{M_\alpha-1}$
depends on $q_{1~i}$ and $p_{1~i}$, then in the Noether transformations law
the order of the highest possible derivative will be smaller by one:
$\mbox{max}_{\alpha}(M_\alpha-2)$. And generally, in an arbitrary case, when
any of coefficients in front of the constraints of $k$-th stage $\phi_\beta^k$
in the Dirac procedure of breeding the constraints depends on $q_{1~i}$ and
$p_{1~i}$ and all the coefficients in front of the constraints
$\phi_\beta^{k+i}(i=1,\cdots,M_\alpha-k)$ are constants, the order of the
highest possible derivative of coordinates in the Noether transformations
law is $M_\alpha-k$.

The order of the highest derivative of $\varepsilon_\alpha (t)$ contained in
the Noether transformations law is equal always to $M_\alpha-1$. Note that
the amount of group parameters $\varepsilon_\alpha$ and $\varepsilon_{s~i}$
are equal to the number of primary constraints of first class.

\section{Example}
We consider the Lagrangian with constraints of first and second
class when the first-class constraints make up a quasi-algebra of the general
form (the restriction \rf{ideal} is not fulfilled). Examples of that sort
for systems only with first-class constraints are described in our previous
works \cite{GSST:tmf2}-\cite{GSST:JP} including also the cases when the
transformation law in the configuration space contains higher (than the first
order) derivatives of coordinates and, therefore, for a canonicity of the
local-symmetry transformations one must extend (by Ostrogradsky) the initial
phase space.

So, consider the Lagrangian
\be \label{exam5:L}
L=\frac{1}{2}{\dot q_1}^2+\frac{1}{2(q_4+q_5)}{\dot q_2}^2
+\frac{1}{2}{\dot q_3}^2+\frac{1}{2}q_2^2+q_3(q_4-q_5).
\ee
Then passing to the Hamiltonian formalism we obtain the generalized momenta
$$p_1=\dot{q}_1, \quad p_2=\frac{\dot q_2}{q_4+q_5}, \quad p_3=\dot{q}_3,
\quad p_4=0, \quad p_5=0$$
and, thus, two primary constraints
\be \label{exam5:phi-prim}
\phi_1^1=p_4, \qquad \phi_2^1=p_5
\ee
and the total Hamiltonian
\be \label{exam5:H_c}
H_T=\frac{1}{2}p_1^2+\frac{1}{2}(q_4+q_5)p_2^2+\frac{1}{2}p_3^2
-\frac{1}{2}q_2^2-q_3(q_4-q_5) + u_1\phi_1^1+u_2\phi_2^1.
\ee
>From the self-consistency conditions of theory we obtain two secondary
constraints
$$
\phi_1^2=-\frac{1}{2}p_2^2+q_3,\qquad \phi_2^2=-\frac{1}{2}p_2^2-q_3,
$$
two tertiary constraints
$$
\phi_1^3=-q_2p_2+p_3,\qquad \phi_2^3=-q_2p_2-p_3
$$
and two quaternary constraints
$$
\phi_1^4=-(q_4+q_5)p_2^2-q_2^2+q_4-q_5,\quad
\phi_2^4=-(q_4+q_5)p_2^2-q_2^2-q_4+q_5.
$$
There do no longer arise constraints, because the conditions of the time
conservation of constraints $\phi_1^4$ and $\phi_2^4$ determine one of the
Lagrangian multipliers. Further one can see for oneself that
~$\mbox{rank}\|\{\phi_\alpha^{m_\alpha},\phi_\beta^{m_\beta}\}\|=4$;
therefore, four constraints are of second class. Now implementing our
procedure of the constraint separation into first and second class, we obtain
the following set of independent constraints:~
the first-class constraints
$$\Phi_1^1=\frac{1}{2}(p_4+p_5),\quad \Phi_1^2=-\frac{1}{2}p_2^2,\quad
\Phi_1^3=-q_2p_2,\quad \Phi_1^4=-(q_4+q_5)p_2^2-q_2^2$$
and the three-linked chain of second-class constraints
$$\Psi_1^1=\frac{1}{2}(p_4-p_5),\quad \Psi_1^2=q_3,\quad \Psi_1^3=p_3, \quad
\Psi_1^4=q_4-q_5.$$
One can see that the first-class constraint $\Phi_1^4$ violates the
condition \rf{ideal}, namely,
\be \label{non-ideal}
\bigl\{\Phi_1^1,\Phi_1^4\bigr\}=-2\Phi_1^2.
\ee
Therefore we shall pass to an equivalent set of constraints by the
transformation \rf{Phi-bar-Phi}:
\be \label{exam5:Phi-bar-Phi}
\tilde{\Phi}_1^{m_1}=C^{m_1 m_1^\p}\Phi_1^{m_1^\p},
\ee
where the matrix ${\bf C}$ is the solution of the equation \rf{C-equation}.
Since from the quantities $f_{\alpha~~\delta~~\gamma}^{1~~m_\delta m_\gamma}$
in \rf{C-equation} the only non-vanishing one is $f_{1~1~1}^{1~4~2}=-2$, the
matrix
\bea \label{exam5:C}
{\bf C}=\left(\begin{array}{cccc} 1 & 0 & 0 & 0 \\ 0 & 1 & 0 & 0
\\ 0 & 0 & 1 & 0 \\ 0 & c & 0 & 1\end{array} \right),
\eea
where $c$ is the solution of the equation~~
$\bigl\{\Phi_1^1,c\bigr\}-2=0$,~~e.g.
\be \label{exam5:c}
c=-2(q_5+q_6),
\ee
can be taken as the particular solution of eq.\rf{C-equation}. Thus we obtain
the desired canonical set of constraints when the condition \rf{ideal} holds
valid~
$(\bigl\{\tilde{\Phi}_1^1,\tilde{\Phi}_1^{m_1}\bigr\}=0,~~ m_1=2,3,4)$:
$$\tilde{\Phi}_1^{m_1^\p}=\Phi_1^{m_1^\p}~~(m_1^\p=1,2,3),\quad
\tilde{\Phi}_1^4=-q_2^2.$$
This provide the fulfilment of the second condition \rf{PB-G-Phi1} of the
conservation of the primary-constraint surface $\Sigma_1$ under the
transformations \rf{q-q^prime}.

Further we seek the generator $G$ in the form \rf{G-Dirac}:
\be \label{exam5:G-Dirac}
G=\eta_1^{k_1}~\Psi_1^{k_1}+\varepsilon_1^{m_1}~\tilde{\Phi}_1^{m_1}, \qquad
k_1=1,\cdots,4,~~~m_1=1,\cdots,4.
\ee
Since in eq.\rf{PB-Psi-H^prime} the only non-vanishing structure functions are
$h_{1~1}^{3~4}=h_{1~1}^{2~3}=h_{1~1}^{2~2}=1$, the system of equations (II.21)
for the coefficients of constraints $\Psi_1^{k_1}$ in eq.\rf{exam5:G-Dirac}
has the form
\bea \label{exam5:eq:eta} \left\{\begin{array}{l}
\dot{\eta}_1^4+\eta_1^3=0,\\\dot{\eta}_1^3+\eta_1^2=0,\\\dot{\eta}_1^2+
\eta_1^1=0.
\end{array}\right.
\eea
Then, taking into account that the first condition \rf{PB-G-Psi1} of the
$\Sigma_1$ conservation under transformations \rf{q-q^prime} gives
~$\eta_1^4=0,$~ we verify on the basis of \rf{exam5:eq:eta} that all
$\eta_1^{k_1}=0$, i.e. the second-class constraints of system
do not contribute to the generator $G$.

As $g_{1~1}^{4~4}=g_{1~1}^{3~3}=g_{1~1}^{4~2}=g_{1~1}^{2~2}=0,~
g_{1~1}^{3~4}=g_{1~1}^{2~3}=g_{1~1}^{1~2}=1$ and
$g_{1~1}^{4~3}=g_{1~1}^{3~2}=2(q_4+q_5)$ in eq.\rf{PB-Phi-H^prime}, the system
of equations \rf{eq:eps} for determining $\varepsilon_1^{m_1}$ becomes
\bea \label{exam5:eq:eps} \left\{\begin{array}{l}
\dot{\varepsilon}_1^4+\varepsilon_1^3=0,\\\dot{\varepsilon}_1^3+
2(q_4+q_5)\varepsilon_1^4+\varepsilon_1^2=0,\\\dot{\varepsilon}_1^2+
2(q_4+q_5)\varepsilon_1^3+\varepsilon_1^1=0.
\end{array}\right.
\eea
Denoting ~$\varepsilon_1^4\equiv \varepsilon$,~ we obtain
\be \label{exam5:eps}
\varepsilon_1^3=-\dot{\varepsilon},~~\varepsilon_1^2=\ddot{\varepsilon}-
2(q_4+q_5)\varepsilon,~~\varepsilon_1^1=-\frac{d}{dt}
\bigl[\ddot{\varepsilon}-2(q_4+q_5)\varepsilon\bigr]+
2(q_4+q_5)\dot{\varepsilon}.
\ee
We see that the quantity $G$ in \rf{exam5:G-Dirac} depends on $\dot{q_4}$ and
$\dot{q_5}$; therefore, for a canonicity of the desired local-symmetry
transformations it is necessary to extend the phase space according to
section 3. It is sufficient to carry out the following extension:
Define the coordinares $\tilde{q}_i~~(i=1,\cdots,7)$:
\be \label{exam5:enlarged-q}
\tilde{q}_i=q_i~~(i=1,\cdots,5) ,\quad
\tilde{q}_6=\dot{q}_4,~~~\tilde{q}_7=\dot{q}_5,
\ee
and their conjugate momenta calculated in accordance with \rf{enlarged-p}
\be
\label{exam5:enlarged-p}
\tilde{p}_i=p_i~~(i=1,\cdots,5),\quad \tilde{p}_6=\tilde{p}_7=0.
\ee
The generalized momenta $\tilde{p}_6$ and $\tilde{p}_7$ are extra primary
constraints of the first class.

In the extended phase space, one should carry out the procedure of
reparametrization of the system of equations \rf{exam5:eq:eps}, although
formally, to obtain the definite form, it is sufficient to express $G$ in the
coordinates of this extended space according to
\rf{exam5:enlarged-q},\rf{exam5:enlarged-p} and \rf{dot-p}:
\bea \label{exam5:enlarged-G}
G&=&\biggl[-\frac{\stackrel{\ldots}{\varepsilon}}{2}+2(\tilde{q}_4+
\tilde{q}_5)\dot{\varepsilon}+(\tilde{q}_6+\tilde{q}_7)~\varepsilon~\biggr]
(\tilde{p}_4+\tilde{p}_5) \nn\\
&&~~+\biggl[-\frac{\ddot{\varepsilon}}{2}+(\tilde{q}_4+
\tilde{q}_5)\varepsilon\biggr]~\tilde{p}_2^2
+\tilde{q}_2(\tilde{p}_2~\dot{\varepsilon}-\tilde{q}_2~\varepsilon).
\eea
It can be seen that the local-symmetry transformations generated by this $G$
\rf{exam5:enlarged-G} are already canonical.

In the $(q,\dot q)$-space the local-symmetry transformations established
with the help of formulas \rf{q-dot-q} have the form
\bea
&&\delta q_1=\delta q_3=0,~~~\delta q_2=-\frac{\dot{q}_2}{q_4+q_5}~
\ddot{\varepsilon}+2\dot{q}_2~\varepsilon+q_2~\dot{\varepsilon}, \nn\\
&&\delta q_4=\delta q_5=-\frac{\stackrel{\ldots}{\varepsilon}}{2}+2(q_4+q_5)~
\dot{\varepsilon}+(\dot q_4+\dot q_5)~\varepsilon,~~
\delta\dot q_i=\frac{d}{dt}~\delta q_i.
\eea
It is easy to verify that under these
transformations
$$ \delta
L=\frac{d}{dt}\biggl\{\frac{\dot{q}_2^2}{(q_4+q_5)^2}
\Bigl[-\frac{\ddot{\varepsilon}}{2}+(q_4+q_5)~\varepsilon~\Bigr]
+q_2^2~\varepsilon~\biggr\}, $$
i.e. the action is quasi-invariant.

\section{Conclusion}

In the framework of the generalized Hamiltonian formalism for dynamical systems
with first- and second-class constraints, we have suggested the method of
constructing the generator of local-symmetry transformations for arbitrary
degenerate Lagrangians both in the phase and configuration space. The general
case is considered including both the violation of the condition \rf{ideal}
(i.e. without restrictions on the algebra of first-class constraints) and the
possibility of the presence of higher derivatives of coordinates in the
local-symmetry transformation law; and the arising problem of canonicity of
transformations in the latter case is solved.

The generator of local-symmetry transformations is derived from the
requirement for them to map the solutions of the Hamiltonian equations of
motion into the solutions of the same equations which must be supplemented by
the demand on the primary-constraint surface $\Sigma_1$ to be conserved under
these transformations. As it is discussed in paper II, the condition of
the $\Sigma_1$ conservation actually is not an additional restriction on the
properties of the local-symmetry transformation generator that naturally
follows from the definition of the symmetry group of the action functional.

We have proved in the general case that that all first-class constraints
generate the local-symmetry transformations (however, the number of the gauge
degrees of freedom equals the number of the primary ones) and second-class
constraints do not contribute to the law of these transformations and do not
generate global transformations in lack of first-class constraints.

The generator of local-symmetry transformations is obtained for degenerate
theories of general form, without restrictions on the algebra of constraints.
We have shown that in this case (these are, e.g., Polyakov's string
\cite{Polyakov} and other model Lagrangians \cite{Cawley},
\cite{Frenkel}-\cite{GSST:JP}) one can always pass to an equivalent set of
constraints, the algebra of which satisfies the condition \rf{ideal}, and,
therefore, now the method of constructing the generator developed for singular
theories of special form in paper II can be applied. In Appendix A, the method
of passing to one of the indicated equivalent sets when all the first-class
primary constraints are momentum variables is given.

The corresponding transformations of local symmetry in the $(q,\dot q)$-space
are determined with the help of formulae \rf{q-dot-q}.

When deriving the local-symmetry transformation generator the employment of
obtained equation system \rf{eq:eps} is important, the solution of which
manifests a mechanism of appearance of higher derivatives of coordinates and
group parameters in the Noether transformation law in the configuration space,
the highest possible order of coordinate derivatives being determined by the
structure of the first-class constraint algebra, and the order of the highest
derivative of group parameters in the transformation law being by unity
smaller than the number of stages in deriving secondary constraints of first
class by the Dirac procedure.

We have shown the obtained local-symmetry transformations to be canonical in
the extended (by Ostrogradsky) phase space where the time derivatives of
coordinates (which have emerged in the transformation law) are taken as
complementary coordinates and the conjugate momenta (defined by the formula of
theories with higher derivatives \cite{Ostr,Gitman,Nesterenko}) are the
initial momenta plus the extra first-class primary constraints (the number of
the latter equals the number of complementary coordinates). In addition,
the dynamics of a system remains to be fixed in the sector of the initial
phase-space variables.

Obtained generator \rf{enl.G} (\rf{G}) satisfies the group property
\rf{enl.G:group}. The amount of group parameters which determine the rank of
quasigroup of these transformations equals the number of primary constraints
of first class.

So, we can state in the general case of theories with first- and second-class
constraints (without restrictions on the constraint algebra) that the
necessary and sufficient condition for a certain quantity $G$ to be the
local-symmetry transformation generator is the representation of $G$ as a
linear combination of all the first-class constraints (and only of them) of
the equivalent set of the special form (when the first-class primary
constraints are the ideal of algebra of all the first-class constraints)
with the coefficients determined by the system of equations \rf{eq:eps}.
Passing to the indicated equivalent set of constraints is always possible, and
the method is presented in this work.
In addition, these are the necessary and sufficient conditions for
\rf{q-q^prime} to be the quasi-invariance transformation of the functional of
action in both the phase and $(q,\dot q)$ space.
It is thereby shown in the general case that the functional of action and the
corresponding Hamiltonian equations of motion are invariant under the same
quasigroup of local-symmetry transformations.

As it is known, gauge-invariant theories belong to the class of degenerate
theories. In this paper, we have shown that the degeneracy of theories with
the first- and second-class constraints in the general case is due to their
quasi-invariance under local-symmetry transformations.

\section*{Acknowledgments}
One of the authors (S.A.G.) thanks the Russian Foundation for Fundamental
Research (Grant N$^{\underline {\circ}}$ 96-01-01223) for support.

\section*{Appendix A}
Here we shall describe the way of passing to, at least, one separated set of
equivalent constraints $\bar \phi_\alpha^{m_\alpha}$ when all the primary
constraints of the first class are momentum variables. We shall consider that
the initial set of first- and second-class constraints are canonical, i.e. the
complete separation of constraints into first- and second-class ones is
already carried out. Then, the formulated problem can be solved by the
iteration procedure provided that we take into account the first-class
primary constraints to make a subalgebra of quasi-algebra of all the
first-class constraints \rf{f}:
$$\{\Phi_\alpha^1,\Phi_\beta^1\}=f_{\alpha~\beta~\gamma}^{1~1~1}\Phi_\gamma^1,
\qquad \alpha,\beta=1,\cdots,F.$$
This relation follows from the stationarity condition for $\Phi_\alpha^1$ and from
from the properties of the canonical set of constraints. The iteration
procedure can be first developed for first-class constraints, and second-class
constraints can be taken into account at last stage. There always exist
canonical transformations of the form \cite{Gitman,Eisenhart}
\bea \label{P-bar}
\bar P_1=\Phi_1^1(q,p),\qquad \{\bar Q_1,\bar P_1\} = 1,\qquad \{\bar Q_\sigma,
\bar P_\tau\} =\delta_{\sigma \tau},\nn\\
\{\bar P_1,\bar P_\tau\} =\{\bar Q_1,\bar P_\tau\} =\{\bar P_1,\bar Q_\tau\} =
\{\bar Q_1,\bar Q_\tau\} = 0,
\eea  $$\sigma,\tau = 2,\cdots,N.$$
(The bar over a letter means the first stage of the iteration procedure.)
All the remaining primary constraints of first class assume the form
$$\varphi_\alpha^1(\bar Q,\bar P)=\Phi_\alpha^1\bigl(q(\bar Q,\bar P),
p(\bar Q,\bar P)\bigr)\Bigr|_{\bar P_1 =0} ,\qquad \alpha= 2,\cdots,F.$$
In view of the transformation being canonical, we can write
$$\{\bar P_1,\varphi_\alpha^1\}=-\frac{\partial\varphi_\alpha^1}{\partial\bar
Q_1}=\bar f_{1~\alpha~\gamma}^{1~1~1}\varphi_\gamma^1,\qquad \alpha,\gamma
\geq 2,$$
with $\varphi_\alpha^1$ having the structure \cite{Gitman}
\be \label{bar:varphi}
\varphi_\alpha^1=\bar E_{\alpha~\gamma}^{1~1}\bar \varphi_\gamma^1,\qquad
\mbox{det} \bar {\bf E}\Bigr|_\Sigma \not= 0,
\ee
and obeying the conditions
$$\frac{\partial\bar \varphi_\gamma^1}{\partial\bar Q_1}=
\frac{\partial\bar \varphi_\gamma^1}{\partial\bar P_1}=0,\qquad \gamma\geq 2.$$
As all the constraints $\bar \varphi_\gamma^1$ do not depend upon $\bar Q_1$
and $\bar P_1$, we perform an analogous procedure for the constraint $\bar
\varphi_2^1$ in the $2N-2$-dimensional subspace $(\bar Q_\sigma,\bar P_\sigma)
(\sigma=2,\cdots,N)$, i.e. without affecting $\bar Q_1$ and $\bar P_1$. Then
the constraints ${\bar {\bar \varphi}}_\alpha^1(\alpha=3,\cdots,F)$ arising
in a formula analogous to formula \rf{bar:varphi} are independent of $\bar
Q_1,\bar P_1$ and ${\bar {\bar Q}}_2,{\bar {\bar P}}_2$. Next, making this
procedure step by step $(F-2)$ times, we finally obtain the first-class
primary constraints to be momenta, and therefore they commute with each other
(final momenta and coordinates will be denoted by $Q_\alpha$ and $P_\alpha$,
respectively, $\alpha=1,\cdots,F$).

All secondary constraints of first class will then assume the form
$$\varphi_\alpha^{m_\alpha}(Q,P)=\varphi_\alpha^{m_\alpha}\Bigl(q(Q,P),p(Q,P)
\Bigr){\Bigr|_{P_\alpha=0}}, \quad \alpha=1,\cdots,F;~ m_\alpha=2,\cdots,M_
\alpha.$$
As the transformations are canonical, we can write
$$\{P_\alpha,\varphi_\beta^{m_\beta}\}=-\frac{\partial\varphi_\beta^{m_\beta}}
{\partial\bar Q_\alpha}=f_{\alpha~~\beta~~\gamma}^{1~~m_\beta m_\gamma}
\varphi_\gamma^{m_\gamma},$$
with $\varphi_\alpha^{m_\alpha}$ having the structure \cite{Gitman}
\be \label{bar:varphi-secondary}
\varphi_\alpha^{m_\alpha}=A_{\alpha~~\beta}^{m_\alpha m_\beta}\widetilde
{\varphi}_\beta^{m_\beta} ,\qquad \mbox{det}{\bf A} {\Bigr|_\Sigma} \not= 0,
\ee
and obeying the conditions
$$\frac{\partial \widetilde {\varphi}_\alpha^{m_\alpha}}{\partial Q_\beta}=
\frac{\partial\widetilde {\varphi}_\alpha^{m_\alpha}}{\partial P_\beta}= 0,
\qquad \alpha,\beta=1,\cdots,F,\quad m_\alpha \geq 2.$$
And, finally, all second-class constraints will be expressed as
$$\psi_{a_i}^{m_{a_i}}(Q,P)=\Psi_{a_i}^{m_{a_i}}\Bigl(q(Q,P),p(Q,P)\Bigr)
{\Bigr|_{P_\alpha=0}}, \quad \alpha=1,\cdots,F;$$
$$i=1,\cdots,n,~a_i=1,\cdots,A_i,~m_{a_i}=1,\cdots,M_{a_i}$$
with all (previously-established in paper I) features of the canonical set
of constraints remaining valid.

The set of constraints thus constructed (primary constraints being momenta and
secondary $\widetilde {\varphi}_\alpha^{m_\alpha}$) satisfies the condition
\rf{ideal} with vanishing right-hand side, i.e. we have derived the searched
set of constraints. Note that $(A^{-1})_{\alpha~~\beta}^{m_\alpha m_\beta}$
in \rf{bar:varphi-secondary} is a solution to the system of equations
\rf{C-equation}.


\begin{thebibliography}{99}
\bibitem{CGS-2} N.P. Chitaia, S.A. Gogilidze and Yu.S. Surovtsev,
``Second-Class Constraints and Local Symmetries,'' Communication of Joint
Institute for Nuclear Research E2-96-234, Dubna, 1996; hep-th/9704085;
a united version of this reference and the present work to be published in
{\it Phys. Rev. D}.
\bibitem{Sugano-Kimura} R. Sugano and T. Kimura, {\it Phys. Rev. D} {\bf 41}
(1990), 1247.
\bibitem{Cabo} A. Cabo and P. Louis-Martinez, {\it Phys. Rev. D} {\bf 42}
(1990), 2726.
\bibitem{Lusanna} L. Lusanna, {\it Riv. Nuovo Cimento} {\bf 14) (1991}, 1.
\bibitem{GSST} S.A. Gogilidze, V.V. Sanadze, Yu.S. Surovtsev and F.G.
Tkebuchava, ``The Theories with Higher Derivatives and Gauge-Transformation
Construction,'' Preprint of Joint Institute for Nuclear Research E2-87-390,
Dubna, 1987; {\it Int. J. Mod. Phys. A} {\bf 47} (1989), 4165.
\bibitem{Bergmann} J.L. Anderson and P.G. Bergmann, {\it Phys. Rev.} {\bf 83}
(1951), 1018.  P.G. Bergmann and J. Goldberg, {\it Phys. Rev.} {\bf 98}
(1955), 531.
\bibitem{Zanelli} M. Henneaux, C. Teitelboim and J. Zanelli, {\it Nucl. Phys.
B} {\bf 332} (1990), 169.
\bibitem{Cabo} A. Cabo and P. Louis-Martinez, {\it Phys. Rev. D} {\bf 42}
(1990), 2726.
\bibitem{Castellani} L. Castellani, {\it Ann. Phys. (N.Y.)} {\bf 143} (1982)
357.
\bibitem{Cawley} R. Cawley, {\it Phys. Rev. Lett.} {\bf 42} (1979), 413;
{\it Phys. Rev. D} {\bf 21} (1980), 2988.
\bibitem{Polyakov} A.M. Polyakov, {\it Phys. Lett. B} {\bf 103} (1981), 207.
\bibitem{Frenkel} A. Frenkel, {\it Phys. Rev. D} {\bf 21} (1980), 2986.
\bibitem{Sugano} R. Sugano and T. Kimura, {\it Prog. Theor. Phys.} {\bf 69}
(1983), 1241.
\bibitem{Gotay} M.J. Gotay, {\it J. Phys. A.: Math. Gen.} {\bf 16} (1983),
L-141.
\bibitem{Gotay-Nester} M.J. Gotay and J.M. Nester, {\it J. Phys. A.: Math.
Gen.} {\bf 17} (1984), 3063.
\bibitem{Stefano} R. Di Stefano, {\it Phys. Rev. D} {\bf 27} (1983), 1758.
\bibitem{Nest-Cherv} V.V. Nesterenko and A.M. Chervyakov, ``Singular
Lagrangians. Classical Dynamics and Quantization,'' Preprint of Joint
Institute for Nuclear Research P2-86-323, Dubna, 1986.
\bibitem{Batlle} C. Batlle, J. Gomis, X. Gr\`acia and J.M. Pons, {\it J. Math.
Phys.} {\bf 30} (1989), 1345.
\bibitem{Gracia-Pons} X. Gr\`acia and J.M. Pons, {\it J. Phys. A: Math. Gen.}
{\bf 25} (1992), 6357.
\bibitem{GSST:tmf2} S.A. Gogilidze, V.V. Sanadze, Yu.S. Surovtsev and
F.G. Tkebuchava, {\it Theor. Math. Phys.} {\bf 102} (1995), 47.
\bibitem{GSST:JP} S.A. Gogilidze, V.V. Sanadze, Yu.S. Surovtsev and
F.G. Tkebuchava, {\it J.Phys. A: Math. Gen.} {\bf 27} (1994), 6509.
\bibitem{GSST:tmf1} S.A. Gogilidze, V.V. Sanadze, Yu.S. Surovtsev and
F.G. Tkebuchava, {\it Theor. Math. Phys.} {\bf 102} (1995), 40.
\bibitem{CGS-1} N.P. Chitaia, S.A. Gogilidze and Yu.S. Surovtsev,
``Constrained Dynamical Systems: Separation of Constraints into First and
Second Classes,'' Preprint of Joint Institute for Nuclear Research E2-96-227,
Dubna, 1996;  hep-th/9704019; to be published in {\it Phys. Rev. D}.
\bibitem{Dirac}  P.A.M. Dirac, {\it Canad. J. Math.} {\bf 2} (1950), 129.
``Lectures on Quantum Mechanics,'' Belfer Graduate School of Science,
Monographs Series, Yeshiva University, New York, 1964.
\bibitem{Smirnov} V.I. Smirnov, ``Course of Higher Mathematics,'' Vol.4,
Part 2, Nauka, Moscow, 1981 (in Russian).
\bibitem{Ostr} M.V. Ostrogradsky, {\it Mem. de l'Acad. Imper. des Sci. de
St-Petersbourg} {\bf 4} (1850), 385.
\bibitem{Gitman} D.M. Gitman and I.V. Tyutin, ``Canonical quantization of
constrained fields,'' Nauka, Moscow, 1986 (in Russian).
\bibitem{Nesterenko}  V.V. Nesterenko, {\it J.Phys. A: Math. Gen.} {\bf 22}
(1989), 1673.
\bibitem{Eisenhart} L.P. Eisenhart, ``Continuous Groups of Transformation,''
Princeton, N.J., Dover, New York, 1961.


\end{thebibliography}
\end{document}